# AI-derived layer-specific OCT biomarkers for classification of geographic atrophy

Yukun Guo[1,2], An-Lun Wu[1,3], Tristan T. Hormel[1], Liqin Gao[1], Min Gao[1,2], Thomas S. Hwang[1], Steven T. Bailey[1], Yali Jia[1,2,*]

[1]*Casey Eye Institute, Oregon Health & Science University, Portland, OR 97239, USA*
[2]*Department of Biomedical Engineering, Oregon Health & Science University, Portland, OR 97239, USA*
[3]*Department of Ophthalmology, Mackay Memorial Hospital, Hsinchu 300044, Taiwan*
*\*jiaya@ohsu.edu*

**Abstract:** Geographic atrophy (GA) is a key biomarker of dry age-related macular degeneration (AMD) traditionally identified through color fundus photography. Hyper-transmission defects (hyperTDs), a feature highly correlated with GA, have recently gained prominence in optical coherence tomography (OCT) research. OCT offers cross-sectional imaging of the retina, leading to the development of the terms complete retinal pigment epithelium and outer retinal atrophy (cRORA) to describe specific patterns of structural degeneration. Within the definitions of cRORA three critical lesions are implicated: inner nuclear layer and outer plexiform layer (INL-OPL) subsidence, ellipsoid zone and retinal pigment epithelium (EZ-RPE) disruption, and hyperTDs. To enable the automated quantification of retinal atrophy progression, we propose an AI-based model that segments INL-OPL subsidence, EZ-RPE disruption, and hyperTDs. Additionally, we developed an algorithm that leverages these segmentation results to distinguish cRORA from hyperTDs in the absence of GA. We evaluated our approach on a comprehensive dataset of eyes with AMD and healthy eyes, achieving mean voxel-level F1-scores of $0.76 \pm 0.13$ (mean $\pm$ standard deviation) for INL-OPL subsidence, $0.64 \pm 0.15$ for EZ-RPE disruption, and $0.69 \pm 0.04$ for hyperTDs. For distinguishing cRORA from hyperTDs, we achieved an average pixel-level F1-score of $0.80 \pm 0.12$ for segment cRORA from hyperTDs. This method demonstrates significant advances in the quantitative analysis of retinal atrophy, offering a promising tool for improved AMD diagnosis and disease progression monitoring.

## 1. Introduction

Geographic atrophy (GA) is a late-stage manifestation of dry age-related macular degeneration (AMD) resulting in significant central vision loss [1,2]. As a critical biomarker of late-stage AMD, GA has attracted considerable interest in AMD research. In color fundus photography (CFP), GA presents as well-demarcated areas of partial or complete depigmentation of the retinal pigment epithelium (RPE), often allowing for visualization of the underlying choroidal vasculature. [3] Since CFP has historically been the primary modality for imaging GA, early research efforts on quantifying GA areas were focused on CFP images. Sunness, J. S., et al. proposed a method for measuring areas of GA in advanced AMD on 30° CFP of the central macula [4]. Due to the low contrast and low resolution of the CFP images, the approach required extensive manual delineation, which limited its practical use.

With the advent of advanced imaging technologies, the study of GA has transitioned from insights gained through a single imaging modality to more advanced, multi-modality research. Fundus autofluorescence (FAF) is also a non-invasive imaging modality that utilizes the fluorescent properties of lipofuscin within the RPE for imaging, allowing it to detect early changes in the RPE and provide better quantification of GA than CFP [5]. The study by Holz et al. demonstrated that the FAF can be used for screening the development of GA due to its ability to image lipofuscin granules in RPE cells [6]. Hu et al. proposed an algorithm based on a supervised pixel classification approach that can successfully automate segmenting GA in

FAF images [7]. However, like CFP, FAF also provides only two-dimensional (2D) images of GA, which limits the ability to examine detailed features of the lesion, particularly changes in the retinal tissue within and surrounding the affected area.

Optical coherence tomography (OCT) generates non-invasive retinal images that, unlike FAF and CFP, provide high-resolution and three-dimensional (3D) images [8] providing layer specific anatomic information about the retina tissue *in vivo*. Several studies have compared and contrasted these three imaging modalities for the assessment of GA. Chen et al. reported that OCT has better visualization of GA than CFP [11]. Velaga et al. studied the correlation between FAF and *en face* OCT in the measurements of GA [12]. In their study, the definition of GA was the area of hyper-transmission defects (hyperTDs) in the choroid as observed on the 2D *en face* projections of OCT. Their result indicated that the *en face* OCT-based GA area measurements are highly repeatable and highly correlated with FAF-based measurements. Hu et al. presented an approach to automatically segment GA in both OCT and FAF images and reported high correlation between GA detection accuracy in the two image modalities [13]. Furthermore, Cleland et al. reported that there are no statistically significant differences in GA measurements using OCT, FAF, CFP, and infrared reflectance imaging [14].

However, unlike CFP and FAF, which provide only 2D images and suffer from low contrast or resolution, OCT offers high-resolution 3D data. With volumetric OCT scans, the changes within the atrophic area are able be visualized in cross-section, enabling more precise quantification of GA extent than using either *en face* images (whether CFP, FAF, or OCT). Wu et al. introduced nascent GA (nGA) [9] that describes features that portend the development of drusen-associated atrophy based on OCT. These hallmark OCT imaged features include the subsidence of the outer plexiform layer and inner nuclear layer (INL-OPL subsidence), and a hypo-reflective wedge-shaped band within the Henle fiber layer. Sadda et al. introduced new categories of degenerative retinal changes, including complete RPE and outer retinal atrophy (cRORA) [10], based on the changes to the outer retina and RPE observed in OCT scans. The term cRORA includes: INL-OPL subsidence, ellipsoid zone (EZ)-RPE disruption, and (hyperTDs) in the choroid. Consequently, several studies have employed OCT to quantify GA [15–20] by defining GA regions as hyperTDs within the choroid. This methodology is also supported by a longitudinal study demonstrating no significant differences between GA areas delineated on FAF and those identified as hyperTDs on OCT [21].

Although existing OCT-based segmentation methods perform well in quantifying hyperTDs, they fall short in accurately measuring key biomarkers essential for defining cRORA, such as INL-OPL subsidence and EZ-RPE disruption. While the current clinical definition of cRORA is sufficiently clear for identification, its quantification remains subjective due to the complexity and variability of these three biomarkers on OCT best visualized with cross-sectional OCT. Therefore, there is a need for an objective, automated approach to reliably quantify the extent and make-up of atrophic regions. To address this, Vente et al. proposed a deep learning algorithm to detect iRORA and cRORA based on the segmentation of hyperTDs, EZ loss, and RPE loss or attenuation; however, their method did not account for INL-OPL subsidence [22].

In this study, we developed a deep learning framework to segment three key biomarkers associated with GA on OCT: INL-OPL subsidence, EZ-RPE disruption, and hyperTDs. Furthermore, we introduced a classification algorithm to distinguish cRORA from hyperTDs based on defined criteria. As the definition of iRORA is limited to an observational level and is very ambiguous for algorithm based identification, it cannot be precisely characterized using these three biomarkers alone. Thus, it was not included in the current analysis. Our approach introduces three main innovations: (1) a convolutional neural network capable of simultaneously segmenting three GA-relevant features in cross-sectional OCT images; (2) a novel classification strategy that utilizes segmentation outputs to accurately differentiate cRORA from hyperTDs in *en face* images; and (3) suggests that hyperTDs alone may serve as a promising biomarker for tracking the progression of advanced AMD.

## 2. Methods

*2.1. Data acquisition*

This study was approved by the Institutional Review Board of Oregon Health & Science University (Portland, OR) and conducted in accordance with the tenets of the Declaration of Helsinki. Written informed consent was obtained from all participants prior to enrollment. Volumetric optical coherence tomography (OCT) data were acquired over the central 6×6-mm macular region using a high-speed, 120-kHz commercial OCT system (SOLIX; Optovue/Visionix, Inc.). At each of the 512 raster scan positions, two repeated B-scans were obtained, with each B-scan comprising 512 A-lines to ensure high spatial resolution and minimizing motion artifacts. Structural OCT volumes were generated by averaging the paired B-scans at each position.

*2.2 Study Inclusion Criteria*

All participants were aged 50 years or older, and a single eye from each participant was imaged to acquire OCT volume data. Two diagnostic groups were included in the study: **Dry AMD group**: Eyes clinically diagnosed with advanced nonexudative (dry) age-related macular degeneration (AMD) exhibiting geographic atrophy (GA), confirmed by multimodal imaging (CFP, OCT) and comprehensive clinical assessment. **Healthy control group**: Eyes (age-matched) with no clinical signs of AMD or other retinal pathologies, verified through detailed ophthalmic examination and imaging. OCT volumes with a signal strength index (SSI) greater than 55 and minimal motion artifacts were included, ensuring suitability for reliable structural analysis.

*2.3. Convolutional Neural Network Design*

In this study, we designed a CNN to segmentate INL-OPL subsidence, EZ-RPE disruption, and hyperTDs in OCT images (Fig. 1). This model follows an encoder-decoder framework with skip connections, inspired by U-Net-like architectures but enhanced with residual learning for improved feature propagation and gradient flow. The encoder consists of a series of residual modules, including both a standard residual module and a residual module with down-sampling (highlighted in blue and orange, respectively). These modules progressively reduce the spatial resolution of the feature maps while increasing the depth, allowing the network to learn hierarchical and abstract representations of the input OCT scan. The decoder mirrors the encoder structure, consisting of up-sampling layers (green blocks) followed by concatenation with the corresponding encoder feature maps (skip connections). This design facilitates the recovery of spatial details lost during down-sampling by reusing high-resolution contextual features from the encoder. Each concatenated feature map is followed by a residual module to refine the segmentation output.

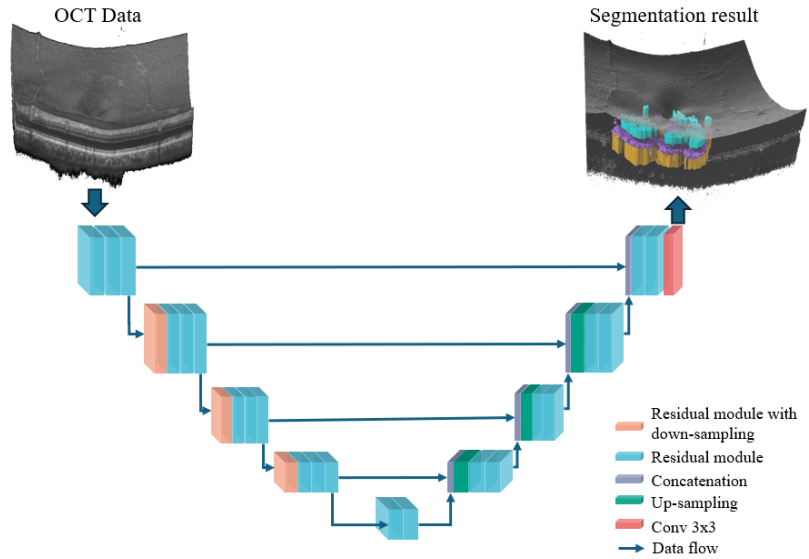

**Figure 1.** Overview of the proposed convolutional neural network architecture for segmenting three retinal biomarkers in OCT images. The model adopts an encoder-decoder structure with residual connections. The encoder consists of residual modules with and without down-sampling to extract hierarchical features, while the decoder uses up-sampling and skip connections to recover spatial resolution. Concatenation of encoder and decoder features enables precise segmentation of key biomarkers, including INL-OPL subsidence, EZ-RPE disruption, and hyper-transmission defects (hyperTDs). INL - inner nuclear layer, OPL - outer plexiform layer, ONL – outer nuclear layer, EZ – ellipsoid zone, RPE – retinal pigment epithelium.

*2.4 Dataset Preprocessing*

The volumetric OCT data was enhanced by averaging adjacent pairs of B-scans to improve the signal-to-noise ratio. Central normalization was applied to standardize the data to zero-mean and unit-variance, thereby improving training stability and convergence speed. The ground truth delineations for the three biomarkers were manually delineated by two experienced graders (Y.G. and M.G.), with reference to retinal layer segmentation[23,24] to ensure anatomical accuracy (Fig. 2). For INL-OPL subsidence and EZ-RPE disruption, the annotations strictly followed the corresponding retinal layer boundaries. The INL-OPL subsidence is defined as the downward shift of both the upper boundary of the INL and the lower boundary of the OPL, where each boundary is displaced by at least 50 percent below its average normal position. EZ–RPE disruption was defined as attenuation or complete loss of reflectivity involving either the EZ or RPE layer on OCT. For the delineation of hyperTDs, the upper boundary was defined at the position of Bruch's membrane, while the lower boundary was set at twice the thickness of the choroid to fully encompass the extent of the hyper-transmission signal.

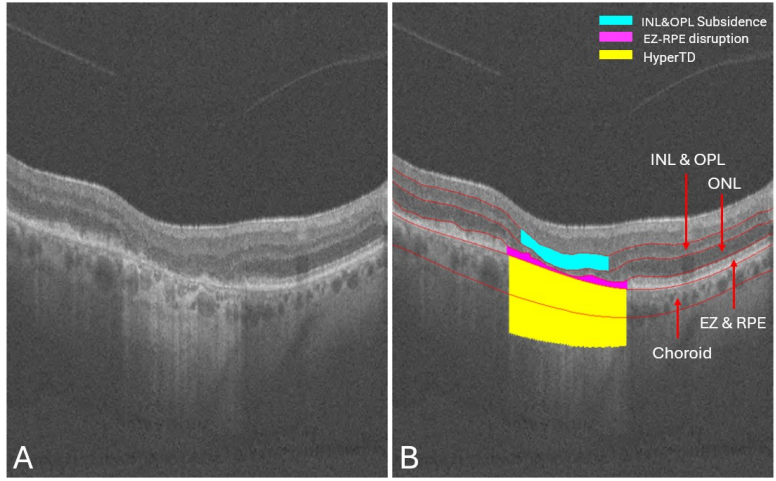

**Figure 2.** Example of biomarker delineation on an OCT B-scan. (A) original OCT image. (B) annotated image showing INL-OPL subsidence (cyan), EZ-RPE disruption (magenta), and hyperTDs (yellow). Red lines represent key retinal layer boundaries between retinal layers used for reference in the annotation process, including the INL - inner nuclear layer, OPL - outer plexiform layer, ONL – outer nuclear layer, EZ – ellipsoid zone, RPE – retinal pigment epithelium, and choroid.

*2.5 Training settings*

To mitigate the effects of class imbalance in the segmentation task, a composite loss function was employed, integrating categorical cross-entropy with Dice coefficient loss. This combined approach enables the model to strike a balance between pixel-wise classification accuracy and region-level overlap, thereby enhancing both local and global segmentation performance. The loss function is formally defined as:

$$L = -\frac{1}{N}\sum_{i=1}^{N}\{Y_i(\log \hat{Y}_i + (1-Y_i)\cdot \log(1-\hat{Y}_i)\} + \left(1 - \sum_{i=1}^{N}\frac{2\hat{Y}_i \cdot Y_i}{\hat{Y}_i + Y_i + \varepsilon}\right) \quad (1)$$

where $N$ denotes the total number of pixels in the input image, $Y_i$ is the ground truth label for pixel $i$, $\hat{Y}_i$ is the predicted probability for the same pixel, and $\varepsilon$ is a small constant added to avoid division by zero.

By incorporating both cross-entropy and Dice loss, the model is encouraged to accurately classify individual pixels while maintaining the structural integrity of the segmented regions. The network was trained using the Adam optimizer with an initial learning rate of 0.001, providing efficient and stable convergence during optimization. The learning rate was reduced by a factor of 10 when the training loss plateaued, defined as a change of less than 0.0001 over three consecutive epochs. To prevent overfitting, an early stopping strategy was employed. The maximum number of training epochs was set at 500, with early stopping triggered if the loss showed minimal improvement (change < 0.0001) over five consecutive epochs. The dataset was split into a training set and test set with a ratio of 4:1. Five-fold cross-evaluation was applied to evaluate the performance of the model.

*2.6 Algorithm for differentiating cRORA from hyperTDs*

The designed algorithm for differentiating cRORA from hyperTDs in *en face* images is based on the segmentation outputs of the three biomarkers from the deep learning model and the definition of cRORA from Classification of Atrophy Meetings [10]. To distinguish cRORA from hyperTDs, the algorithm first projects the volumetric segmentations of the three

biomarkers—INL-OPL subsidence, EZ-RPE disruption, and hyperTDs—on to 2D *en face* maps. It then identifies regions where either INL-OPL subsidence or EZ-RPE disruption is present and checks for spatial overlap between these disruptions and the hyperTDs areas. If the area of overlap between the structural disruptions and the hyperTDs exceeds a predefined threshold (set at 0.05 mm²), the region is classified as cRORA. Any remaining hyperTDs that do not meet this overlap criterion are labeled as residual hyperTDs. The detailed procedure of the algorithm is presented in **Table 1**. It is important to note that the threshold T is derived from the area of an equivalent circle corresponding to the original 250 μm diameter cutoff used to distinguish iRORA from cRORA on B-scans. This area-based threshold is adopted in the algorithm to provide a more stable and consistent differentiation in the *en face* projection.

Table 1. Procedure for Differentiating cRORA from hyperTDs

| | Algorithm 1. Differentiating cRORA from hyperTDs |
|---|---|
| 1 | Procedure Diff-cRORA-HyperTDs (INL-OPL_volume, EZ-RPE_volume, hyperTDs_volume) |
| 2 | INLOPL, EZRPE, hyperTDs ← Projection2D (INLOPL_volume, EZRPE_volume, hyperTDs_volume) |
| 3 | INLOPL_EZRPE ← IPONL ∪ EZRPE |
| 4 | INLOPL_EZRPE _in_hyperTDs ← INLOPL_EZRPE ∩ hyperTDs |
| 5 | cRORA ← Area (INLOPL_EZRPE _in_hyperTDs) > T* |
| 6 | hyperTD_residual ← hyperTDs ∩ ¬cRORA |
| 7 | End Procedure |

*: T is set to 0.05mm²

## 3. Results

### 3.1 Study population

In total 96 participants, 71 with AMD and 25 healthy controls were collected from a clinical AMD study.

### 3.2 Segmentation accuracy

Following training, the performance of the proposed model was evaluated on a held-out test set to assess its generalizability to unseen data (Table 2). In volumetric (voxel-wise) evaluation, it demonstrated balanced performance across all three biomarkers, with high specificity and moderate-to-high sensitivity and F1-scores, indicating effective detection with minimal false positives. When the same evaluation was performed on projected *en face* images, segmentation accuracy further improved, showing consistently higher sensitivities and F1-scores while maintaining strong specificity. Moreover, leveraging the *en face* biomarker maps enables reliable differentiation of cRORA from hyperTDs, achieving both high sensitivity and specificity.

Table 2. Quantification of segmentation accuracy (Mean ± Standard Deviation) on three biomarkers (Five-fold cross-validation)

| | INL & OPL Subsidence | EZ-RPE disruption | hyperTDs | cRORA |
|---|---|---|---|---|
| Volumetric accuracy in voxel | | | | |
| Sensitivity | 0.76±0.15 | 0.75±0.12 | 0.89±0.08 | - |
| Specificity | 0.99±0.00 | 0.99±0.00 | 0.99±0.01 | - |
| F1-score | 0.76±0.13 | 0.64±0.15 | 0.69±0.04 | - |
| En Face accuracy in pixel | | | | |
| Sensitivity | 0.90±0.10 | 0.90±0.08 | 0.98±0.02 | 0.98±0.03 |
| Specificity | 0.99±0.01 | 0.96±0.03 | 0.98±0.02 | 0.98±0.01 |
| F1-score | 0.82±0.11 | 0.74±0.18 | 0.86±0.08 | 0.80±0.12 |

From the representative case shown in Figure 3, the proposed model can identify isolated small areas of hyperTDs in the absence of both INL & OPL subsidence and EZ-RPE disruption, demonstrating its sensitivity to subtle signal changes that may not meet criteria for cRORA (Fig. 3 A, B, C, D. orange arrowhead). At another region (Fig 3. A, B, E, F, green arrowhead) highlights early GA features, where EZ-RPE disruption is present without accompanying INL & OPL subsidence or hyperTDs. The model accurately segmented these isolated features. In a region exhibiting prominent drusen (blue arrowhead, Panels A, B, G, H), the model correctly identified the co-localization of all three biomarkers—OPL-INL subsidence, EZ-RPE disruption, and hyperTDs—around or overlying the drusenoid deposits.

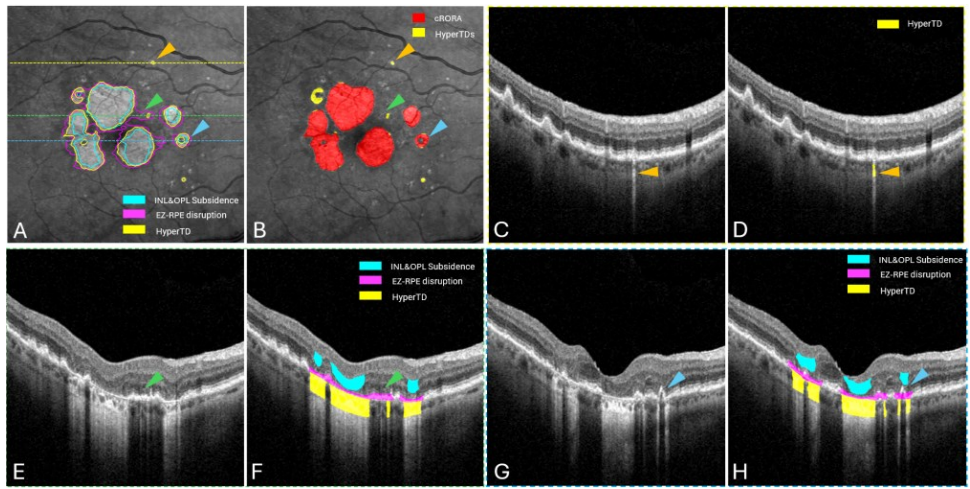

**Figure 3.** An Example of biomarker segmentation and cRORA classification on OCT. (A) *En face* projection image of cRORA with borders of INL & OPL subsidence (cyan), borders ofEZ-RPE disruption (magenta), and borders of hyperTDs ellow). (B) En face projection of classified cRORA regions (shaded red) and hyperTD (shaded yellow) in absence of INL&OPL subsidence and EZ-RPE disruption.The yellow dashed line indicates the position of B-scans in C and D, and the green dashed line indicates the position of the B-scan in E and F, the blue dashed line indicates the position of B-scan in G and H. (C, D) B-scans corresponding to the orange arrowhead showing isolated hyperTDs without INL & OPL subsidence or EZ-RPE disruption. (E, F) B-scans at the green arrowhead illustrate an early atrophic lesion with only EZ-RPE disruption present. (G, H) B-scans at the blue arrowhead demonstrate a drusen-associated region with all three features co-localized.

### 3.3 Longitudinal analysis of cRORA and hyperTDs

We validated the proposed algorithm using a longitudinal OCT dataset from a patient with advanced dry AMD, covering a 46-month follow-up. As shown in Figure 4, the en face OCT images (A1–H1) and segmented overlays illustrate the spatial progression of hyperTDs (yellow) and cRORA (red). The combined overlays (A4–H4) highlight the evolving relationship between these biomarkers over time. Quantitative measurements show a steady increase in lesion areas (Fig. 5). Both cRORA and hyperTDs show continuous enlargement. The residual hyperTDs, representing non-overlapping regions, remained small and stable, with many eventually converting to cRORA.

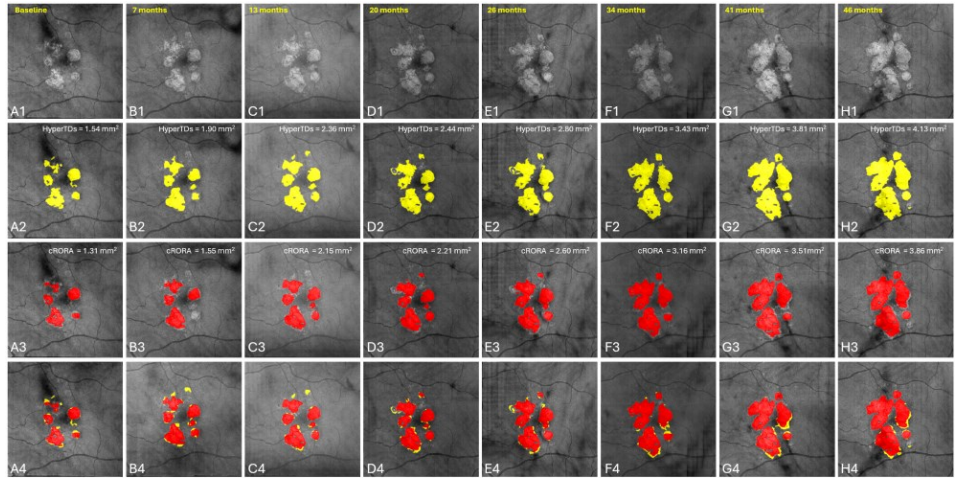

**Figure 4.** Longitudinal tracking of GA progression in a patient with advanced dry AMD over 46 months. (A1–H1) En face OCT projection images from baseline to 46 months. (A2–H2) Segmented hyperTDs shown in yellow, with area measurements indicating progressive expansion. (A3–H3) Segmented cRORA regions shown in red, reflecting the enlargement of atrophic areas over time. (A4–H4) Combined overlays of cRORA (red) and residual hyperTDs (yellow), illustrating the temporal and spatial dynamics of GA progression.

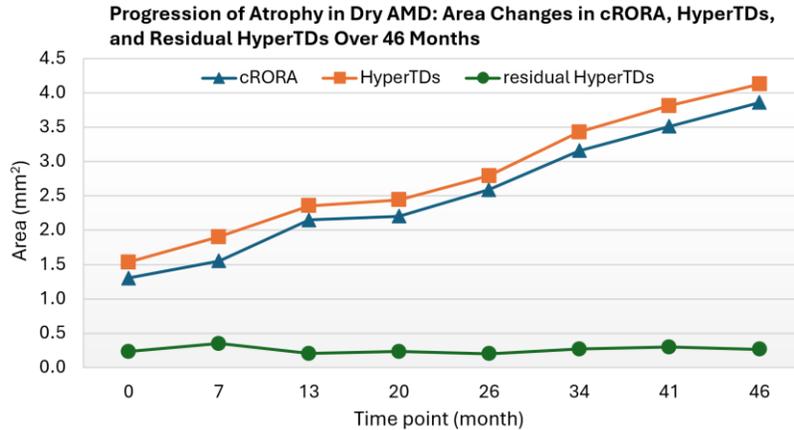

**Figure 5.** Dynamic changes of cRORA, hyperTDs, and residual hyperTDs over 46 months. The line chart shows the progression of cRORA (blue triangles), total hyperTDs (orange squares), and residual hyperTDs (green circles) across eight time points (baseline to 46 months).

## 4. Discussion

In this study, we developed and validated a deep learning-based framework for segmenting and classifying key structural features associated with GA in OCT imaging, including INL-OPL subsidence, EZ-RPE disruption, and hyperTDs. By leveraging these features, we designed a classification algorithm to distinguish cRORA from hyperTDs based on the criteria reported in a previous study [10]. This pipeline was trained using manually annotated OCT data and demonstrated strong performance on the test datasets.

As summarized in Table 2, the segmentation network demonstrates solid F1-score across all three features in volumetric analysis. When evaluated with *en face* images, the F1-scores improved further, indicating more accurate and reliable delineation. In addition, the model achieved a strong F1-score when distinguishing cRORA from hyperTDs, highlighting its potential for clinically meaningful classification. Segmentations in cross-sectional and *en face* images showed high consistency with the visualization of the GA region (Fig 3). Furthermore, longitudinal application of the algorithm to follow-up scans in a patient with dry AMD (Fig. 4) revealed progressive increases in both cRORA and hyperTDs, demonstrating the system's effectiveness in capturing GA progression over time. Additionally,, the residual hyperTDs may serve as a potential biomarker for predicting the progression rate of GA (Fig. 5). The ability to quantify and distinguish between evolving lesion components supports their utility in both clinical trials and routine disease monitoring.

Unlike previous studies that primarily perform direct segmentation of GA regions on en face OCT images [25] or rely on a single feature in cross-sectional OCT scans [26], our approach introduces a more comprehensive analysis of the INL, photoreceptors, and RPE by automatically segmenting three distinct anatomical features associated with GA: INL-OPL subsidence, EZ-RPE disruption, and hyperTDs. Previous *en face*–based methods often depend on intensity thresholds or projection maps, which can be influenced by noise, image contrast, or projection artifacts, leading to subjective or less anatomically precise boundaries. Similarly, cross-sectional methods focusing on a single feature may overlook subtle structural interactions among retinal layers. Notably, our definition of INL-OPL subsidence differs slightly from that in previous studies [27,28]. In our approach, all three features are defined strictly according to the anatomic boundaries of the retinal layers. This more objective criterion improves the specificity of subsidence detection and allows for more accurate identification of outer retinal changes.

In this study, the definition of hyperTDs differs from that used in previous works [29–31], where the hyper-transmission signal was extracted from a fixed sub-RPE slab spanning 65 to 400 μm beneath the RPE. To enhance anatomical relevance and reduce the influence of large choroidal vessels, our hyperTDs layer extended from Bruch's membrane to a depth equivalent to twice the choroidal thickness. This structure-aware definition provides a more consistent anatomically adaptive and consistent basis for signal extraction, improving differentiation from other hyperreflective features within the choroid. By enlarging the detection region, the new definition also mitigates the effects of large choroidal vessels and shadow artifacts. Signal reduction caused by these features is compensated by the inclusion of a greater proportion of the surrounding tissue, leading to more stable hyperTDs detection. With this more reliable signal characterization, our algorithm can detect early changes in the EZ and RPE by identifying subtle or isolated hyperTD region, potentially indicating regions at higher risk of progression to GA. Additionally, because our model detects hyperTDs directly from cross-sectional OCT rather than *en face* images, the segmented hyperTD regions may contain focal regions of hypo-transmission secondary signal blockage, due hyper-reflective features such as calcified (refractile) drusen [32]. Although these holes may not have immediate clinical significance, they could serve as a potential feature for monitoring GA progression.

In this study, we also designed an algorithm to distinguish cRORA from hyperTDs in *en face* images by integrating three key biomarkers. Our algorithm defines cRORA as any region larger than 0.05 mm² where hyperTDs co-localize with either INL-OPL subsidence or EZ-RPE disruption. This strategy aligns with the structural features described by a clinical consortium[cite] which introduced clearer and more objective criteria. Compared to the original definition of cRORA, our more sophisticated analysis of the INL, photoreceptors and RPE may provide more relevant prognostic information. In this study, we were able to demonstrate reliable and automated quantification. This refinement may help reduce ambiguity in borderline cases and improve consistency in longitudinal monitoring.

Despite its contributions, this study has several limitations. First, our layer specific anatomical deterioration model did not account for incomplete RORA (iRORA) because iRORA definition is moreambiguous and imprecise term. It is possible that hyperTD on the edge of GA alone may be adequate for predicting regions of GA extension. Second, the proposed model does not distinguish between the disruption of the EZ and the RPE, as combining them improved segmentation performance. However, this separation could be considered in future work by incorporating precise layer segmentation between the EZ and RPE. Lastly, the clinical relevance of the newly defined hyperTDs was not fully explored in this study and will be investigated as the next step.

## 5. Conclusions

In this study, we proposed a deep learning-based framework for the automated segmentation of three key OCT features related to GA: INL-OPL subsidence, EZ-RPE disruption, and hyperTDs. Using these segmented features, we developed an algorithm to distinguish cRORA from hyperTDs based on clinically established diagnostic criteria. The framework showed strong performance in both cross-sectional and longitudinal evaluations, successfully tracking GA progression over time. By focusing on anatomically defined features, our approach offers a reliable and interpretable tool to support GA both diagnostic and prognostic studies.

**Founding**

This work was supported by grants from National Institutes of Health (R01 EY 036429, R01 EY035410, R01 EY024544, R01 EY027833, R01 EY031394, R43EY036781, P30 EY010572, T32 EY023211, UL1TR002369); the Jennie P. Weeks Endowed Fund; the Malcolm M. Marquis, MD Endowed Fund for Innovation; Unrestricted Departmental Funding Grant and Dr. H. James and Carole Free Catalyst Award from Research to Prevent Blindness (New York, NY); Edward N. & Della L. Thome Memorial Foundation Award, and the Bright Focus Foundation (G2020168, M20230081).

**Disclosures**

Yukun Guo: Optovue/Visionix (P), Genentech/Roche (P); Ifocus Imaging (P), Tristan T. Hormel: Ifocus Imaging (I), Tristan T. Hormel: Ifocus Imaging (I), Steven Baily: Visionix/Optovue (F), Yali Jia: Optovue/Visionix (P, R), Genentech/Roche (P, R, F), Ifocus Imaging (I, P), Optos (P), Boeringer Ingelheim (C), Kugler (R)

**Data availability**

Data underlying the results presented in this paper are not publicly available at this time but may be obtained from the authors upon reasonable request.

**Reference**

1. A. Gupta, R. Bansal, A. Sharma, and A. Kapil, "Macular Degeneration, Geographic Atrophy, and Inherited Retinal Disorders," in *Ophthalmic Signs in Practice of Medicine* (Springer Nature Singapore, 2023), pp. 351–396.
2. R. Sacconi, E. Corbelli, L. Querques, F. Bandello, and G. Querques, "A Review of Current and Future Management of Geographic Atrophy," Ophthalmol Ther **6**(1), 69–77 (2017).
3. Age-Related Eye Disease Study Research Group, "The Age-Related Eye Disease Study system for classifying age-related macular degeneration from stereoscopic color fundus photographs: the Age-Related Eye Disease Study Report Number 6.," Am J Ophthalmol **132**(5), 668–81 (2001).
4. J. S. Sunness, N. M. Bressler, Y. Tian, J. Alexander, and C. A. Applegate, "Measuring geographic atrophy in advanced age-related macular degeneration.," Invest Ophthalmol Vis Sci **40**(8), 1761–9 (1999).
5. A. A. Khanifar, D. E. Lederer, J. H. Ghodasra, S. S. Stinnett, J. J. Lee, S. W. Cousins, and S. Bearelly, "Comparison of color fundus photographs and fundus autofluorescence images in measuring geographic atrophy area.," Retina **32**(9), 1884–91 (2012).
6. F. G. Holz, C. Bellman, S. Staudt, F. Schütt, and H. E. Völcker, "Fundus autofluorescence and development of geographic atrophy in age-related macular degeneration.," Invest Ophthalmol Vis Sci **42**(5), 1051–6 (2001).
7. Z. Hu, G. G. Medioni, M. Hernandez, and S. R. Sadda, "Automated segmentation of geographic atrophy in fundus autofluorescence images using supervised pixel classification," Journal of Medical Imaging **2**(1), 014501 (2015).

8. D. Huang, E. A. Swanson, C. P. Lin, J. S. Schuman, W. G. Stinson, W. Chang, M. R. Hee, T. Flotte, K. Gregory, and C. A. Puliafito, "Optical coherence tomography.," Science **254**(5035), 1178–81 (1991).
9. Z. Wu, C. D. Luu, L. N. Ayton, J. K. Goh, L. M. Lucci, W. C. Hubbard, J. L. Hageman, G. S. Hageman, and R. H. Guymer, "Optical coherence tomography-defined changes preceding the development of drusen-associated atrophy in age-related macular degeneration.," Ophthalmology **121**(12), 2415–22 (2014).
10. S. R. Sadda, R. Guymer, F. G. Holz, S. Schmitz-Valckenberg, C. A. Curcio, A. C. Bird, B. A. Blodi, F. Bottoni, U. Chakravarthy, E. Y. Chew, K. Csaky, R. P. Danis, M. Fleckenstein, K. B. Freund, J. Grunwald, C. B. Hoyng, G. J. Jaffe, S. Liakopoulos, J. M. Monés, D. Pauleikhoff, P. J. Rosenfeld, D. Sarraf, R. F. Spaide, R. Tadayoni, A. Tufail, S. Wolf, and G. Staurenghi, "Consensus Definition for Atrophy Associated with Age-Related Macular Degeneration on OCT: Classification of Atrophy Report 3.," Ophthalmology **125**(4), 537–548 (2018).
11. Q. Chen, T. Leng, S. Niu, J. Shi, L. de Sisternes, and D. L. Rubin, "A false color fusion strategy for drusen and geographic atrophy visualization in optical coherence tomography images.," Retina **34**(12), 2346–58 (2014).
12. S. B. Velaga, M. G. Nittala, A. Hariri, and S. R. Sadda, "Correlation between Fundus Autofluorescence and En Face OCT Measurements of Geographic Atrophy.," Ophthalmol Retina **6**(8), 676–683 (2022).
13. Z. Hu, G. G. Medioni, M. Hernandez, A. Hariri, X. Wu, and S. R. Sadda, "Segmentation of the geographic atrophy in spectral-domain optical coherence tomography and fundus autofluorescence images.," Invest Ophthalmol Vis Sci **54**(13), 8375–83 (2013).
14. S. C. Cleland, S. M. Konda, R. P. Danis, Y. Huang, D. J. Myers, B. A. Blodi, and A. Domalpally, "Quantification of Geographic Atrophy Using Spectral Domain OCT in Age-Related Macular Degeneration.," Ophthalmol Retina **5**(1), 41–48 (2021).
15. Q. Chen, L. de Sisternes, T. Leng, L. Zheng, L. Kutzscher, and D. L. Rubin, "Semi-automatic geographic atrophy segmentation for SD-OCT images," Biomed Opt Express **4**(12), 2729 (2013).
16. A. Elsawy, T. D. L. Keenan, Q. Chen, X. Shi, A. T. Thavikulwat, S. Bhandari, E. Y. Chew, and Z. Lu, "Deep-GA-Net for Accurate and Explainable Detection of Geographic Atrophy on OCT Scans," Ophthalmology Science **3**(4), 100311 (2023).
17. Z. Chu, L. Wang, X. Zhou, Y. Shi, Y. Cheng, R. Laiginhas, H. Zhou, M. Shen, Q. Zhang, L. de Sisternes, A. Y. Lee, G. Gregori, P. J. Rosenfeld, and R. K. Wang, "Automatic geographic atrophy segmentation using optical attenuation in OCT scans with deep learning," Biomed Opt Express **13**(3), 1328 (2022).
18. Y. Guo, T. T. Hormel, M. Gao, S. T. Bailey, and Y. Jia, "Geographic Atrophy Segmentation on Optical Coherence Tomography," Invest Ophthalmol Vis Sci **65**(9), PB0091–PB0091 (2024).
19. Q. S. You, J. Wang, Y. Guo, C. J. Flaxel, T. S. Hwang, D. Huang, Y. Jia, and S. T. Bailey, "Detection of Reduced Retinal Vessel Density in Eyes with Geographic Atrophy Secondary to Age-Related Macular Degeneration Using Projection-Resolved Optical Coherence Tomography Angiography.," Am J Ophthalmol **209**, 206–212 (2020).
20. S. Niu, L. de Sisternes, Q. Chen, D. L. Rubin, and T. Leng, "Fully Automated Prediction of Geographic Atrophy Growth Using Quantitative Spectral-Domain Optical Coherence Tomography Biomarkers.," Ophthalmology **123**(8), 1737–1750 (2016).
21. C. Simader, R. G. Sayegh, A. Montuoro, M. Azhary, A. L. Koth, M. Baratsits, S. Sacu, C. Prünte, D. P. Kreil, and U. Schmidt-Erfurth, "A longitudinal comparison of spectral-domain optical coherence tomography and fundus autofluorescence in geographic atrophy.," Am J Ophthalmol **158**(3), 557–66.e1 (2014).
22. C. de Vente, P. Valmaggia, C. B. Hoyng, F. G. Holz, M. M. Islam, C. C. W. Klaver, C. J. F. Boon, S. Schmitz-Valckenberg, A. Tufail, M. Saßmannshausen, C. I. Sánchez, and MACUSTAR Consortium, "Generalizable Deep Learning for the Detection of Incomplete and Complete Retinal Pigment Epithelium and Outer Retinal Atrophy: A MACUSTAR Report.," Transl Vis Sci Technol **13**(9), 11 (2024).
23. Y. Guo, T. T. Hormel, S. Pi, X. Wei, M. Gao, J. C. Morrison, and Y. Jia, "An end-to-end network for segmenting the vasculature of three retinal capillary plexuses from OCT angiographic volumes," Biomed Opt Express **12**(8), 4889 (2021).
24. Y. Guo, A. Camino, M. Zhang, J. Wang, D. Huang, T. Hwang, and Y. Jia, "Automated segmentation of retinal layer boundaries and capillary plexuses in wide-field optical coherence tomographic angiography," Biomed Opt Express **9**(9), 4429 (2018).

25. V. Pramil, L. de Sisternes, L. Omlor, W. Lewis, H. Sheikh, Z. Chu, N. Manivannan, M. Durbin, R. K. Wang, P. J. Rosenfeld, M. Shen, R. Guymer, M. C. Liang, G. Gregori, and N. K. Waheed, "A Deep Learning Model for Automated Segmentation of Geographic Atrophy Imaged Using Swept-Source OCT.," Ophthalmol Retina **7**(2), 127–141 (2023).
26. G. Kalra, H. Cetin, J. Whitney, S. Yordi, Y. Cakir, C. McConville, V. Whitmore, M. Bonnay, L. Lunasco, A. Sassine, K. Borisiak, D. Cohen, J. Reese, S. K. Srivastava, and Justis. P. Ehlers, "Machine Learning-Based Automated Detection and Quantification of Geographic Atrophy and Hypertransmission Defects Using Spectral Domain Optical Coherence Tomography," J Pers Med **13**(1), 37 (2022).
27. Z. Wu, E. K. Glover, E. E. Gee, L. A. B. Hodgson, and R. H. Guymer, "Functional Evaluation of Retinal Pigment Epithelium and Outer Retinal Atrophy by High-Density Targeted Microperimetry Testing," Ophthalmology Science **4**(2), 100425 (2024).
28. G. Aresta, T. Araujo, G. S. Reiter, J. Mai, S. Riedl, C. Grechenig, R. H. Guymer, Z. Wu, U. Schmidt-Erfurth, and H. Bogunovic, "Deep Neural Networks for Automated Outer Plexiform Layer Subsidence Detection on Retinal OCT of Patients With Intermediate AMD," (2024).
29. K. B. Schaal, P. J. Rosenfeld, G. Gregori, Z. Yehoshua, and W. J. Feuer, "Anatomic Clinical Trial Endpoints for Nonexudative Age-Related Macular Degeneration.," Ophthalmology **123**(5), 1060–79 (2016).
30. R. Laiginhas, Y. Shi, M. Shen, X. Jiang, W. Feuer, G. Gregori, and P. J. Rosenfeld, "Persistent Hypertransmission Defects Detected on En Face Swept Source Optical Computed Tomography Images Predict the Formation of Geographic Atrophy in Age-Related Macular Degeneration," Am J Ophthalmol **237**, 58–70 (2022).
31. Z. Chu, L. Wang, X. Zhou, Y. Shi, Y. Cheng, R. Laiginhas, H. Zhou, M. Shen, Q. Zhang, L. de Sisternes, A. Y. Lee, G. Gregori, P. J. Rosenfeld, and R. K. Wang, "Automatic geographic atrophy segmentation using optical attenuation in OCT scans with deep learning," Biomed Opt Express **13**(3), 1328 (2022).
32. R. F. Spaide, "PATHWAYS TO GEOGRAPHIC ATROPHY IN NONNEOVASCULAR AGE-RELATED MACULAR DEGENERATION.," Retina **44**(10), 1655–1665 (2024).